\documentclass[aps,pra,preprint,showpacs,preprintnumbers,amsmath,amssymb,footinbib]{revtex4}
\usepackage{graphicx}
\usepackage{bm}
\usepackage{hyperref}
\newcommand{\beq}{\begin{equation}}
\newcommand{\eeq}{\end{equation}}
\newcommand{\bea}{\begin{eqnarray}}
\newcommand{\eea}{\end{eqnarray}}
\newcommand{\mqd}{m^{de}}
\newcommand{\tde}{T_{de}}
\newcommand{\td}{T_d}
\newcommand{\bfq}{{\bf q}}
\newcommand{\calv}{{\cal V}}
\begin{document}

\preprint{BNL-97210-2012-JA, RBRC-948}

\title{Critical endpoint for deconfinement in matrix and other effective
models}
\author{Kouji Kashiwa}
\email{kashiwa@ribf.riken.jp}
\affiliation{RIKEN/BNL, Brookhaven National Laboratory, 
Upton, NY 11973}
\author{Robert D. Pisarski}
\email{pisarski@bnl.gov}
\affiliation{
Department of Physics, Brookhaven National Laboratory, 
Upton, NY 11973}
\affiliation{RIKEN/BNL, Brookhaven National Laboratory, 
Upton, NY 11973}
\author{Vladimir V. Skokov}
\email{vskokov@quark.phy.bnl.gov}
\affiliation{
Department of Physics, Brookhaven National Laboratory, 
Upton, NY 11973}
\begin{abstract}
We consider the position of the deconfining critical endpoint,
where the first order transition for deconfinement is washed out
by the presence of massive, dynamical quarks.
We use an effective matrix model, employed previously to analyze
the transition in the pure glue theory.
{\it If} the parameters of the pure glue theory are unaffected by
the presence of dynamical quarks, and if the quarks only
contribute perturbatively, then
for three colors and three degenerate quark flavors this quark mass
is very heavy, $\mqd \sim 2.5$~GeV, while the critical
temperature, $\tde$, barely changes, 
$\sim 1 \%$ below that in the 
pure glue theory.  The location of the deconfining critical endpoint 
is a sensitive test to differentiate between
effective models. 
For example, models with a logarithmic potential for the Polyakov loop
give much smaller values of the quark mass, $\mqd \sim 1$~GeV, and
a large shift in $\tde \sim 10\%$ lower than that in the pure glue theory.
\end{abstract}
\maketitle

The nature of the phase transitions in QCD are of interest for a variety of
reasons.  In theory, 
all in-equilibrium thermodynamic quantities can be computed from
first principles using numerical simulations on the lattice.  In practice,
and especially if one is interested in comparing to experiment,
it is useful to have effective models.  These allow one to
compute quantities near thermal equilibrium,
such as transport coefficients, which are much more difficult
to extract from the lattice.

For a pure glue theory with three colors and no dynamical quarks,
the deconfining phase transition is of first order.
This follows from the global $Z(3)$ symmetry in the pure glue theory
\cite{Svetitsky:1982gs}.
Adding dynamical quarks acts like a background $Z(3)$ field, and so tends
to weaken the first order transition.  As the quark mass decreases, 
it is possible to reach
a deconfining critical endpoint, at a mass $\mqd$,
where the deconfining transition is of second order.  

In this paper we consider the properties of the deconfining critical endpoint
in effective models.
This was done before in a zero parameter matrix model 
by Meisinger, Miller, and Ogilvie \cite{Meisinger:2001cq, Meisinger:2001fi}.
In this paper we perform the computation
in one \cite{Dumitru:2010mj} and two \cite{Dumitru:2012}
parameter matrix models.  This is a useful exercise, 
since the solution of the zero parameter model 
does not agree with lattice data on the interaction measure of the pure
$SU(3)$ glue theory. In contrast, the coefficients of the one and two parameter
models are tuned to give increasingly good agreement with the lattice results
\cite{Dumitru:2010mj, Dumitru:2012}.

For these matrix models, and for models based
upon polynomials of the Polyakov loop 
\cite{Pisarski:2000eq, Scavenius:2002ru, Dumitru:2003cf, Ratti:2005jh}, 
we find a heavy $\mqd$.
For three colors and three degenerate quark flavors,
$\mqd$ is about twice as heavy as the charm quark mass,
$\mqd \sim 2.5$~GeV for the matrix model, 
and $\sim 3.5$~GeV for a polynomial loop model.  
This is in sharp contrast to models based on a potential motivated
by the Vandermonde
determinant, which involve the logarithm of the Polyakov loop
\cite{Fukushima:2003fw,Roessner:2006xn,Hell:2009by,Horvatic:2010md,Hell:2011ic}.
They give a quark mass that is lighter than the charm quark, $\mqd \sim 1$~GeV.

Besides the value of $\mqd$, effective models can give detailed
information about the properties of the deconfining critical endpoint.
One obvious parameter is the temperature at which it occurs, $\tde$.
If $\td$ is the temperature of the deconfining phase transition in the
pure glue theory, matrix and polynomial loop models find very small shifts
in the temperature,
$\tde \sim 0.995 \, \td$. The shift in logarithmic loop models is
much larger, $\tde \sim 0.9 \, \td$.  

We also present results for the interaction measure of the theory,
which exhibits characteristic differences between the different models.
For three degenerate flavors, in the matrix
and polynomial loop models, the interaction measure exhibits two peaks,
one near $\td$, and one near $3 \td$.  This is not seen in logarithmic
loop models.  

There are lattice simulations \cite{DeTar:2009ef,Petreczky:2012rq}
which address this problem.
These give a result which is close to the value in a logarithmic
loop model, $\mqd \sim 1.4$~GeV 
\cite{Meisinger:1995qr, Alexandrou:1998wv, Karsch:2001ya}.
They are not current, however.
There are recent results from the WHOT 
collaboration \cite{Saito:2011fs}, but they do not present
an estimate for $\mqd$, nor of $\td$.  
Recently, the Wuppertal-Budapest Collaboration has published numerical 
results on the QCD equation of state with $2+1$ light flavors and a
dynamical charm quark~\cite{Borsanyi:2012vn}.  
Thus they could also address the question of the deconfining critical 
endpoint with relative ease.

At the outset, we note that in our effective model,
any dimensional parameter is a pure number times a common mass scale,
which we chose at $\td$. That is, our model determines
the dimensionless ratios, $\mqd/\td$ and $\td/\td$.  To illustrate
the physics, when we quote quantities such as $\mqd$, we
uniformly assume that $\td = 270$~MeV, but stress that this is an
assumption.

\section{Effective matrix models}

\subsection{Model for a pure glue theory}
\label{pure_glue_model}

We wish to model the region near the deconfining transition temperature.
In this region, which has been termed 
the ``semi'' quark gluon plasma (QGP), the expectation value of
the Polyakov loop is less than unity.  
In a matrix model 
\cite{Meisinger:2001cq, Dumitru:2010mj,Dumitru:2012},
to represent this we expand the time like component of the vector potential 
about a constant value,
\beq
A_0^{i j} = \frac{2 \pi T}{g} \; q_i \; \delta^{i j} \; .
\label{fund_ansatz}
\eeq
For now we consider the general case of $SU(N)$, where
indices in the fundamental representation run from
$i,j=1, 2, \ldots, N$.  The matrix ${A_0}$ is an element of the $SU(N)$ Lie
algebra, so that $\sum_i q_i = 0$.

In this background field, in the fundamental
representation the Wilson line $\bf L$ and
the Polyakov loop $\ell$ are
\beq
{\bf L} = {\rm e}^{ 2 \pi i \, {\bf q} } \;\;\; ; \;\;\;
\ell = \frac{1}{N} \; {\rm tr} \; {\bf L} \; .
\eeq

At one loop order, in this background field the potential for $q$ is
\beq
\calv_{pert}^{gl}(\bfq) \; = \; - \; 
\frac{(N^2 -1) \pi^2}{45} \; T^4 \; + \; {2\pi^2\over 3} \; T^4 \; 
\sum_{i,j=1}^{N} V_2(q_i - q_j) \; ,
\label{pert_gl_pot}
\eeq
where
\beq
V_2(x) = x^2(1-|x|)^2 \;\; , \;\; -1 \leq x \leq 1 \; .
\eeq
$V_2(x)$ is periodic in $x$,
$x \rightarrow x + 1$.  This can be understood more generally.
While the Wilson line is gauge variant, its eigenvalues,
$\exp(2 \pi i q_i)$, are gauge invariant.  

In the pure glue theory, the potential is also invariant under $Z(N)$
transformations.  A $Z(N)$ transformation is given, {\it e.g.}, 
by shifting $q_i \rightarrow q_i + 1/N$
for $i = 1\ldots(N-1)$, and $q_N \rightarrow - (N-1)/N$, under which
${\bf L} \rightarrow \exp(2 \pi i/N) \, {\bf L}$.  The fact that  the potential
is invariant under this transformation is elementary,
since the differences $|q_i - q_j|$ either vanish or equal one, which
by periodicity is equivalent to zero.

In the matrix model, to drive the transition to confinement, 
one adds a non-perturbative term~\cite{Dumitru:2010mj}
\beq
\calv_{non}^{gl}(\bfq) \; = \; - \; {4\pi^2 \over 3} \;
T^2 \, \td^2 \; \left(
c_1\; \sum_{i,j=1}^{N}  V_1(q_i - q_j)+ 
c_2\; \sum_{i,j=1}^{N}  V_2(q_i - q_j) 
%c_1 \; V_1(\bfq)\, + \, c_2\; V_2(\bfq) 
+ \frac{(N^2 -1)}{60} \; c_3 \right) \; .
\label{vtotnpt}
\eeq
where
\beq
V_1(x) = x (1-|x|) \;\; , \;\; -1 \leq x \leq 1 \; .
\eeq
Again, the potential $V_1(x)$ is periodic in $x \rightarrow x + 1$;
it is also invariant under $Z(N)$ transformations.

\subsection{Inclusion of dynamical quarks}
\label{sec:quark}

In the background field of Eq. (\ref{fund_ansatz}), at one loop order
gluons generate the potential in Eq. (\ref{pert_gl_pot}).  When
dynamical quarks are added then, certainly the first thing to do is
to compute the analogous contribution which they make to the $q$-dependent
potential.

This has been computed in Ref. \cite{Meisinger:2001cq}, but we
include it here for completeness.
For quarks of mass $m$, the one loop potential is given by
\begin{align}
\ln \mathrm{det} ( \gamma^\mu \partial_\mu + q \delta^{\mu 4} + i m)
&=
2 \ln \mathrm{det} 
\Bigl[ \Bigl( \partial_0 + 2 \pi T q \Bigr)^2 
            - {\vec \partial}^{\, 2} + m^2\Bigr],
\end{align}
As fermions, in the Matsubara formalism, 
the frequencies are odd multiples of $\pi T$.

For simplicity we denote the background field simply by $q$.
This is equivalent to an imaginary chemical potential for the quark.
This makes it easy to do the integral.  
Instead of summing over the frequencies, we can just use the standard 
expression for the 
thermodynamic potential of a free fermion gas, replacing the chemical
potential $\mu$ by $2 \pi i q$.  Thus we need to compute
\begin{equation}
- \, 2 \, T \int \frac{d^3 p}{(2\pi)^3}
  \Bigl[ \ln \left( 1 + e^{- E_\mathrm{p}/T - 2 \pi i q }\right)  + 
\ln \left(1 + e^{- E_\mathrm{p}/T +  2\pi i q } \right)\Bigr],
\end{equation}
where $E_\mathrm{p} = \sqrt{\vec{p}^{\, 2} + m^2}$.  We then expand each
logarithm in a power series,
\begin{equation}
\ln \Bigl[ 1 + e^{-E_\mathrm{p}/T - 2 \pi i q }
 \Bigr]
= e^{-E_\mathrm{p}/T - 2 \pi i q }
 - \frac{1}{2} \; e^{-2 E_\mathrm{p}/T - 4 \pi i q  } + \cdots
\end{equation}
where the first term is the usual Boltzmann term.  

This leaves the integral over momenta.  The angular integral is trivial,
leaving only the integral over $p$:
\begin{equation}
\int_0^\infty p^2 e^{- E_\mathrm{p}/T} dp
= - \frac{\partial}{\partial \beta}
   \int_0^\infty \frac{p^2}{E_\mathrm{p}} \; e^{-\beta E_\mathrm{p}} dp
= - \frac{\partial}{\partial \beta}
   \int_1^\infty m^2 \sqrt{y^2-1} \; e^{-\beta \, m \, y} dy
\end{equation}
$$
= - \frac{\partial}{\partial \beta}
   \Bigl[ \frac{m}{\beta} K_1(\beta m) \Bigr]
= m^2 \, T \; K_2\left(\frac{m}{T}\right) \; ,
$$
where $K_\nu$ is the modified Bessel function of the second kind, 
and $\beta=1/T$.

The modified Bessel function of the first kind is given by
\begin{equation}
I_{-\nu}(x) - I_{\nu}(x)
= \frac{ \Gamma \Bigl(\frac{1}{2}-\nu \Bigr) 
   \sin(2\nu\pi)}
        {\pi^{{3}/{2}}}
\; \Bigl(\frac{x}{2}\Bigr)^\nu \;
   \int_1^\infty (y^2-1)^{\nu-\frac{1}{2}} e^{-xy} dy,
\end{equation}
where $\Gamma(x)$ is the usual Gamma function.
The Bessel function of the second kind is
\begin{equation}
K_\nu (x) = \frac{\pi}{2} \;
\frac{I_{-\nu}(x) - I_{\nu}(x)} {\sin(\nu\pi)} \; .
\end{equation}

Putting all of the factors together, and summing over different colors,
for a single flavor of massive quark
its contribution to the potential is given by
\begin{equation}
{\cal V}_\mathrm{pert}^{qk}({\bf q})
= \frac{2 \, m^2 \, T^2}{\pi^2} 
\; \sum_{n=1}^\infty
\frac{(-1)^n}{n^2} 
\; K_2 \left(\frac{n \, m}{T} \right) \; 
\sum_{i = 1}^N \; \cos (2 \pi n q_i) \; .
\label{quark_potential}
\end{equation}
The dependence upon the $q$'s can be rewritten more generally as
\beq
\sum_{i = 1}^N \; \cos (2 \pi n q_i)  \; = \;
\frac{ 
{\rm tr} \; {\bf L}^n +
{\rm tr} \; \left({\bf L}^\dagger\right)^n }{2}\; .
\eeq

\subsection{Predictions for matrix model}

We now concentrate on the results for three colors.  
For the pure glue theory, by a $Z(3)$ rotation we can assume that
the Polyakov loop is real.  With dynamical quarks, this remains true
if all of the quark masses are real and the quark chemical potential 
vanishes.  We then
take a path in $q$-space as
\beq
{\bf q} = \frac{1}{3} \left(r-1, 0, 1-r\right) 
\eeq
We have chosen to introduce $r$ because the confined vacuum is then
given by $r = 0$; the perturbative vacuum is $r =1$.

For three colors, assuming that the Polyakov loop
is real, uniquely specifies the $q_i$.  This is not true for four or
more colors \cite{Dumitru:2012}.  We also have computed the position
of the deconfining critical endpoint for this model
in the limit of an infinite number of colors \cite{Pisarski:2012}, and find,
somewhat surprisingly, that the parameters 
for the deconfining critical endpoint are similar to that for three colors.

In writing the potential, it is convenient to introduce a dimensionless
temperature, rescaled by 
$T_d$,
\beq
t = \frac{T}{T_d} \; .
\eeq
The gluon potential is a sum of two terms,
\beq
{\cal V}^{gl}(r,t) =
{\cal V}^{gl}_{pert}(r,t) +
{\cal V}^{gl}_{non}(r,t) 
= \frac{8 \pi^2}{45} \; T_d^4 \; t^2 \; (t^2 - c_2) \;
\left[ {\cal W}_{gl}(r,t) + {\cal W}_{gl}^0(t) \right]\;.
\eeq
The first term is a simple quartic potential in $r$,
\beq
{\cal W}_{gl}(r,t) = - z(t) \; r^2 - \frac{10}{9} \; r^3
+ \frac{5}{3} \; r^4 \; ,
\eeq
where
\beq
z(t) = \frac{5}{3} - \frac{50}{27} \; 
\left( \frac{1 - c_2}{t^2 - c_2}\right) \; .
\eeq
Here we have fixed the constant $c_1$ by requiring 
that the transition occurs at $T_d$; this results in the relation
\beq
c_1 = \frac{50}{27} \; \left(1-c_2 \right) \; .
\label{cond_c1}
\eeq

The second constant, $c_3$, is fixed by requiring that the pressure
vanish at $T_d$, which gives
the condition
\beq
c_3 = 1 + c_1 - \frac{10}{9} \left(1 - c_2\right) \; . 
\label{cond_c3}
\eeq
This
leaves a term independent of $r$ in the potential,
\beq
{\cal W}_{gl}^0(t) = \frac{1}{9} \; 
\left( \frac{t^2 - 1}{t^2 - c_2} \right) \; .
\label{zero_point_termA}
\eeq
The overall constant for $SU(N)$ is $1/N^2$, which is $1/9$ for $N=3$.  
The basic justification of this matrix model is an expansion in large $N$,
so this constant term manifestly represents a correction of $\sim 1/N^2$.

For three colors, the model of Ref. \cite{Meisinger:2001cq}
takes $c_2 = 0$.  The values of $c_1$ and $c_3$ are then fixed by
Eqs. (\ref{cond_c1}) and (\ref{cond_c3}), so this model has
no free parameters.  When $c_2 = 0$, though,
the peak in the interaction measure, $(e-3p)/T^4$ ($e$ is the
energy density and $p$ the pressure),
is much broader than found from lattice simulations
\cite{Dumitru:2010mj,Dumitru:2012}.

We next consider the one parameter model of Ref.
\cite{Dumitru:2010mj}.  This finds a good fit to the interaction measure,
for values
\begin{equation}
c_1 = 0.31537,~~~~ c_2 =  0.8297,~~~~ c_3 = 1.12615 \; .
\label{parameters}
\end{equation}

The one parameter model gives a good fit to the interaction measure
overall, but has problems very close to $T_d$; see, {\it e.g.},
Fig. (3) of Ref. \cite{Dumitru:2012}.  
In particular,
the value of the latent heat for the one parameter model is
much smaller than found on the lattice.
To avoid this discrepancy, we consider 
the two parameter model of Ref. \cite{Dumitru:2012}.  
In this model, $c_1$ and $c_2$ are constants, as before,
but now $c_3$ is a function of temperature,
\beq
c_3(t) = c_3(\infty) + \frac{c_3(1) - c_3(\infty)}{t^2} \; .
\eeq
Because $T^2 c_3$ enters into the potential, this is equivalent to
a MIT ``bag'' constant, 
\beq
B = \frac{8 \pi^2}{45} \left[ c_3(1) - c_3(\infty) \right] \td^4 \; .
\eeq
Fixing the transition to occur at $\td$
relates $c_1$ and $c_2$ as in Eq. (\ref{cond_c1}).
Requiring that the pressure vanish in the confined phase
fixes $c_3(1)$ to have the value of Eq. (\ref{cond_c3}).
That leaves two free parameters, 
$c_2$ and $c_3(\infty)$.  An optimal fit was found to be given by
\begin{equation}
c_1 = 0.830,~~~~ c_2 =  0.5517,~~~~ c_3(1) = 1.332,~~~~
c_3(\infty) = 0.95 \; . 
\label{parametersB}
\end{equation}
The corresponding MIT bag constant is $(244 \; {\rm MeV})^4$.

The $r$-dependent terms in the potential 
are unchanged.  The only change in the potential is the term
independent of $r$, which becomes
\beq
{\cal W}_{gl}^0(0,t) = \frac{1}{9} + 
\left( \frac{-c_1 -c_2 + c_3(t)}{t^2 - c_2} \right) \; .
\eeq
When $c_3(t)$ is a constant, this reduces to the previous form in 
Eq. (\ref{zero_point_termA}).

The addition of dynamical quarks behaves as expected.  
The pure glue theory is invariant under $Z(3)$ transformations, so that
the confining vacuum, where $r = 0$, is always a stationary point of the
potential.  

With quarks, the $Z(3)$ symmetry is lost.  This is clear, 
as the terms in the quark potential, 
Eq. (\ref{quark_potential}), involves
\begin{equation}
\; {\rm tr} \; {\bf L}^n \; = \;
1 + 2 \; \cos\left( \frac{2 \pi n (1-r)}{3}\right) \; .
\end{equation}
While these quantities all vanish in the confined vacuum, what matters
is the equation of motion for $r$, and it is easy to see that any
derivative with respect to $r$ is nonzero when $r = 0$.  Thus the
presence of dynamical quarks removes the $Z(3)$ symmetry of the pure
glue theory, and acts like a background $Z(3)$ field.

Including the quark term, a nonzero expectation value is generated for
any Polyakov loop, ${\rm tr} \, {\bf L}^n$, at any nonzero temperature.
There is a problem with the $1/N^2$ term in the gluon potential, 
Eq. (\ref{zero_point_termA}), though.  While small relative to the
other gluon terms for $T > T_d$, with quarks, there is an expectation
for the Polyakov loop(s), and so a pressure, below $T_d$.  The
term in Eq. (\ref{zero_point_termA}) tends to give negative pressures
below $T_d$, which is manifestly unphysical.

Such a negative pressure is not always present.  For example, in models
like this we could consider the limit in which both the number of colors, $N$,
and the number of quark flavors, $N_f$, are large.  Then for temperatures
below $T_d$ in the pure glue theory, the pressure is large, $\sim N^2$
or $N N_f$, which are both of the same order.

To avoid this problem, for the one parameter model
we simply modify the gluon potential as
\beq
{\cal W}_{gl}^0(t) \rightarrow
{\cal W}_{gl}^0(r,t) = \frac{1}{9} \; 
\left( \frac{t^2 - 1}{t^2 - c_2} \right) \; r^2 \; .
\label{zero_point_termB}
\eeq
That is, to suppress the contribution of ${\cal W}_{gl}^0$, we promote
it to a function of $r$, that is constructed to vanish when $r = 0$.
We admit that this is an {\it ad hoc} procedure.  In principle, one 
could envisage matching 
onto a hadron resonance model in the low temperature phase. 

This modification changes the properties of the deconfining transition
in the pure gauge theory, but these are very small.  In particular,
the transition temperature $\td$ does not shift, since like
${\cal W}_{gl}^0(t)$, ${\cal W}_{gl}^0(r,t)$ vanishes at $\td$, $t=1$.
We have also checked that the change in the parameters for the deconfining
critical endpoint are small.  The main modification is that with
this improvement, the pressure is always non-negative, even below $T_d$.

Similarly, for the two parameter model, by hand we change the
potential as
\beq
{\cal W}_{gl}^0(0,t) \rightarrow
{\cal W}_{gl}^0(r,t) = \left[\frac{1}{9} + 
\left( \frac{-c_1 -c_2 + c_3(t)}{t^2 - c_2} \right)\right] r^2  \; .
\eeq
Besides multiplying by $r^2$, we also investigated multiplying by
$\ell^2$; the results obtained were very similar.  This is because
in the end we are looking for a regime where the pressure is small, anyway.

To compute, we make the most minimal assumptions possible.
Notably, we {\it assume} that the temperature which enters into the pure
glue potential, $T_d$, remains unchanged by the addition of quarks.
Ultimately, the validity of this assumption will be tested by the
comparison of our model to the results of lattice simulations.

We also assume that the {\it only} 
non-perturbative terms are those of the pure glue
theory.  This actually is forced upon us by the $Z(N)$ structure
of the theory.  While quarks break the $Z(N)$ symmetry, numerical
lattice simulations with three colors find that, at least for up to
three light flavors, that the breaking of $Z(3)$ is relatively mild.

This implies that any terms in the non-perturbative potential should
have the symmetry of the pure glue theory, and {\it not} that of the
theory with dynamical quarks.  We have considered the modification of
the theory with terms such as 
$\sim T^2\,  T_d^2 \, {\rm tr} \, {\bf L}^n$, characteristic of
quarks.  These terms break the $Z(3)$ symmetry, and drastically
change the expectation value of the loop, even above $T_d$.  
We have checked that if such terms are present, that their numerical
coefficients must be {\it very} small.  

Lastly, we take the parameter $c_2$ to be that of the pure glue theory.
This is really a rather mild assumption; the important point is that
we do not allow the mass scale $T_d$ to itself change as we add
dynamical quarks.  We also take the values of the other 
constants in the non-perturbative potential,
$c_1$ and $c_3$, Eq. (\ref{parameters}),
from the pure glue theory \cite{Dumitru:2012}.  

In short, we assume that confinement is driven by the transition in the
pure glue theory.  Upon adding quarks, the modification to the theory
is entirely through their perturbative contribution to the $q$-dependent
potential.  

Notably, this means that we assume that the transition temperature
the pure glue theory, $\td$, remains unchanged after the addition of quarks.
Of course adding quarks will shift the physical transition in the theory,
which is why the temperature for the deconfining critical endpoint, $\tde$,
will differ from $\td$.  

While these assumptions are rather strong, they do
allow us to make unique predictions for the theory with
dynamical quarks from the pure glue model.

\section{Other Models}

Two types of the Polyakov loop effective potential  have been
widely used. The first is a potential which is a 
polynomial in the Polyakov loop 
\cite{Pisarski:2000eq, Scavenius:2002ru, Dumitru:2003cf, Ratti:2005jh}:
\begin{equation}
\frac{{\cal V}_{\text{poly}}}{T^4} = -\frac{b_2(T)}{2}\ell^* \ell -
\frac{b_3}{6}
\left[\ell^3 + (\ell^*)^3\right]+\frac{b_4}{4}(\ell^* \ell)^2 \; .
\label{eq:pot_poly}
\end{equation}
The mass term is convoluted,
\begin{equation}
b_2(T) =  a_0 + a_1\frac{\td}{T} + a_2\left( \frac{\td}{T} \right)^2 +
a_3 \left( \frac{\td}{T} \right)^3.
\end{equation}
The coefficients are determined by fitting the equation of state and the
expectation value of the Polyakov loop to lattice data of pure gauge theory
\cite{Boyd:1996bx,Kaczmarek:2002mc} in Ref.~\cite{Ratti:2005jh}. 

The second is motivated by the strong coupling expansion.
Following Fukushima, one uses
a logarithmic potential ~\cite{Fukushima:2003fw,Roessner:2006xn,Hell:2009by,Horvatic:2010md,Hell:2011ic}:
\beq
\frac{{\cal V}_{\text{log}}}{T^4} = -\frac{a(T)}{2}\ell^*
\ell+b(T) \;\ln\left\{1-6\ell^* \ell 
+4\left[\ell^3 + (\ell^*)^3\right] -3(\ell^* \ell)^2\right\} \; .
\label{pot_log}
\eeq
Again, the coefficients are involved functions of temperature,
\begin{equation}
a(T)= a_0 + a_1\frac{\td}{T} + a_2 \left( \frac{\td}{T}
\right)^2,\quad 
b(T) = b_3\left(\frac{\td}{T}\right)^{3}.
\end{equation}
The above parameterization for the temperature dependence was introduced 
by R\"{o}{\ss}ner, Ratti, and Weise \cite{Roessner:2006xn},
with the constants determined
by fitting lattice data to the pure $SU(3)$ theory.

For a single flavor and three colors, the 
quark contribution to the thermodynamic potential for the Polyakov loop
models is
\beq
 {\cal V}_\mathrm{pert}^{qk}({ \ell, \ell^* })
  = -2 T \int \frac{d^3p}{(2\pi)^3}
\left\{ 
\ln[1 + 3(\ell+\ell^* e^{-\beta
E_p})e^{-\beta E_p} + e^{-3\beta E_p}] 
+ {\rm c.c.}
\right\} \; . 
\label{eq:thermo_pot} 
\eeq
We assume that there is no chemical potential for the quarks, so that
$\ell^*=\ell$. In this case, Eq.~(\ref{eq:thermo_pot}) can be 
rewritten similarly 
to Eq.~(\ref{quark_potential}) 
\begin{equation}
{\cal V}_\mathrm{pert}^{qk}(\ell)
= \frac{2 \, m^2 \, T^2}{\pi^2} 
\; \sum_{n=1}^\infty
\frac{(-1)^n}{n^2} 
\; K_2 \left(\frac{n \, m}{T} \right) \; 
\left[ 1 + 2 \; T_n \left( \frac{3\ell -1}{2} \right)\right]
\; ,
\label{quark_potential_loop}
\end{equation}
where $T_n(x)$ is the $n$-th Chebyshev polynomial, 
$T_n(x) = \cos(n \arccos(x))$.

\begin{table}[!t]
\caption{Parameters in the polynomial potential \eqref{eq:pot_poly}.}
\label{tbl:poly_pot}
\begin{ruledtabular}
\begin{tabular}{ccccccc}
$\td$[MeV]&$a_0$&$a_1$&$a_2$&$a_3$&$b_3$&$b_4$ \\ \hline
$270$&$6.75$&$-1.95$&$2.625$&$-7.44$&$0.75$&$7.5$ \\
\end{tabular}
\end{ruledtabular}
\caption{Parameters in the logarithmic potential \eqref{pot_log}.}
\label{tbl:log_pot}
\begin{ruledtabular}
\begin{tabular}[t]{ccccc}
$\td$[MeV]&$a_0$&$a_1$&$a_2$&$b_3$ \\ \hline 270&3.51&$-2.47$&$15.22$&$-1.75$ \\
\end{tabular}
\end{ruledtabular}
\end{table} 

\section{Results}

\begin{figure}
\includegraphics[scale=0.3]{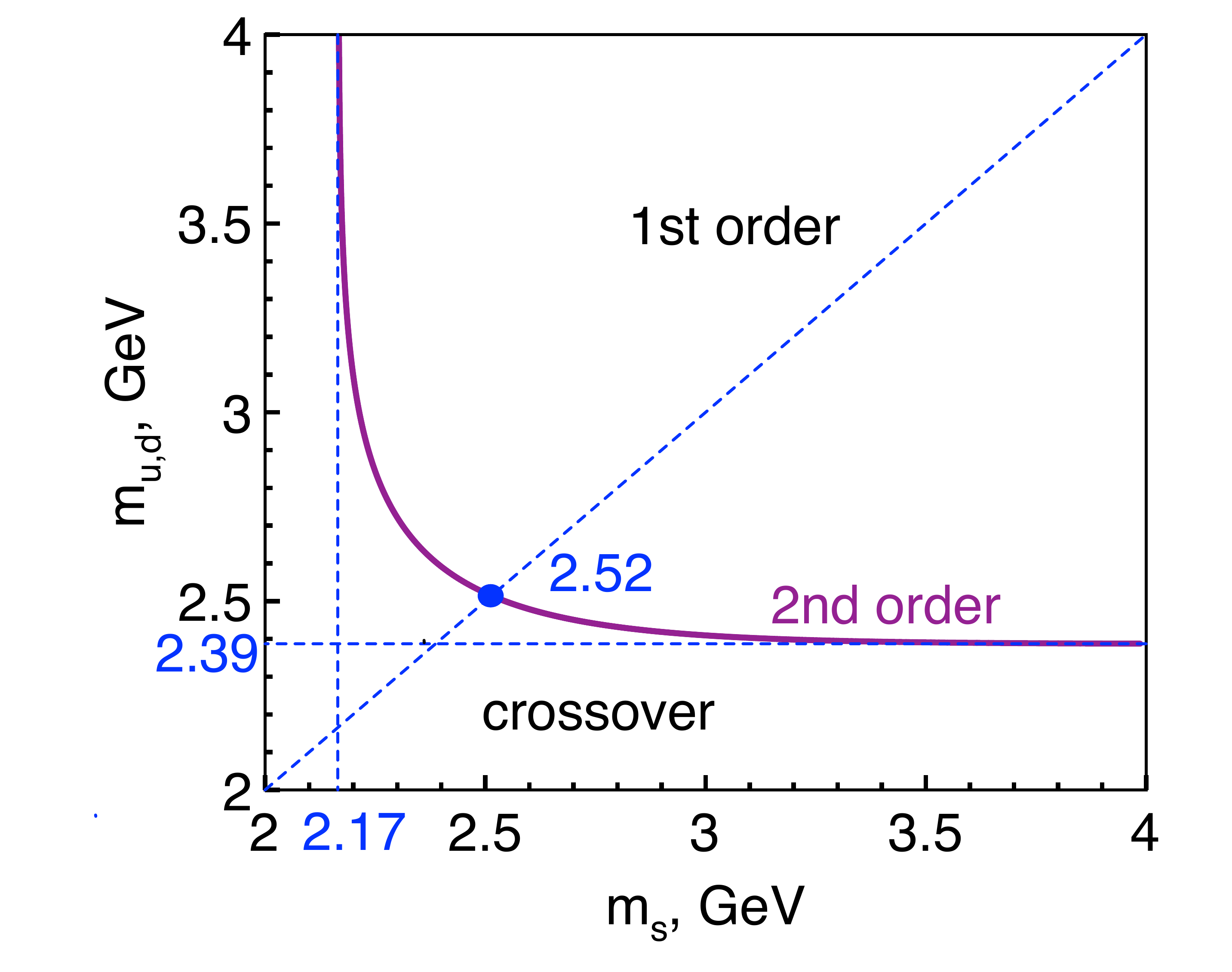}
\caption{
The phase diagram for the deconfining phase transition in the matrix
model without a bag constant, assuming $\td \sim 270$~GeV.
}
\label{fig:TdeM}
\end{figure}

Most of the results can be understood by considering 
Eq.~(\ref{quark_potential}) in the non-relativistic limit, when
$m \gg T$.  Then the 
contribution of a single flavor of quark to the potential is
\beq
{\cal V}_\mathrm{pert}^{qk}({\bf q}) \approx
- \; \frac{\sqrt{2}}{\pi^{3/2}} \;
T^{5/2} \; m^{3/2} {\rm e}^{- m/T} \; {\rm Re} \; {\rm tr} \; {\bf L} \; .
\label{non_rel}
\eeq
We find that for a matrix model, and those with a polynomial
potential, that this is a good approximation.  

\begin{figure}
\includegraphics[scale=0.3]{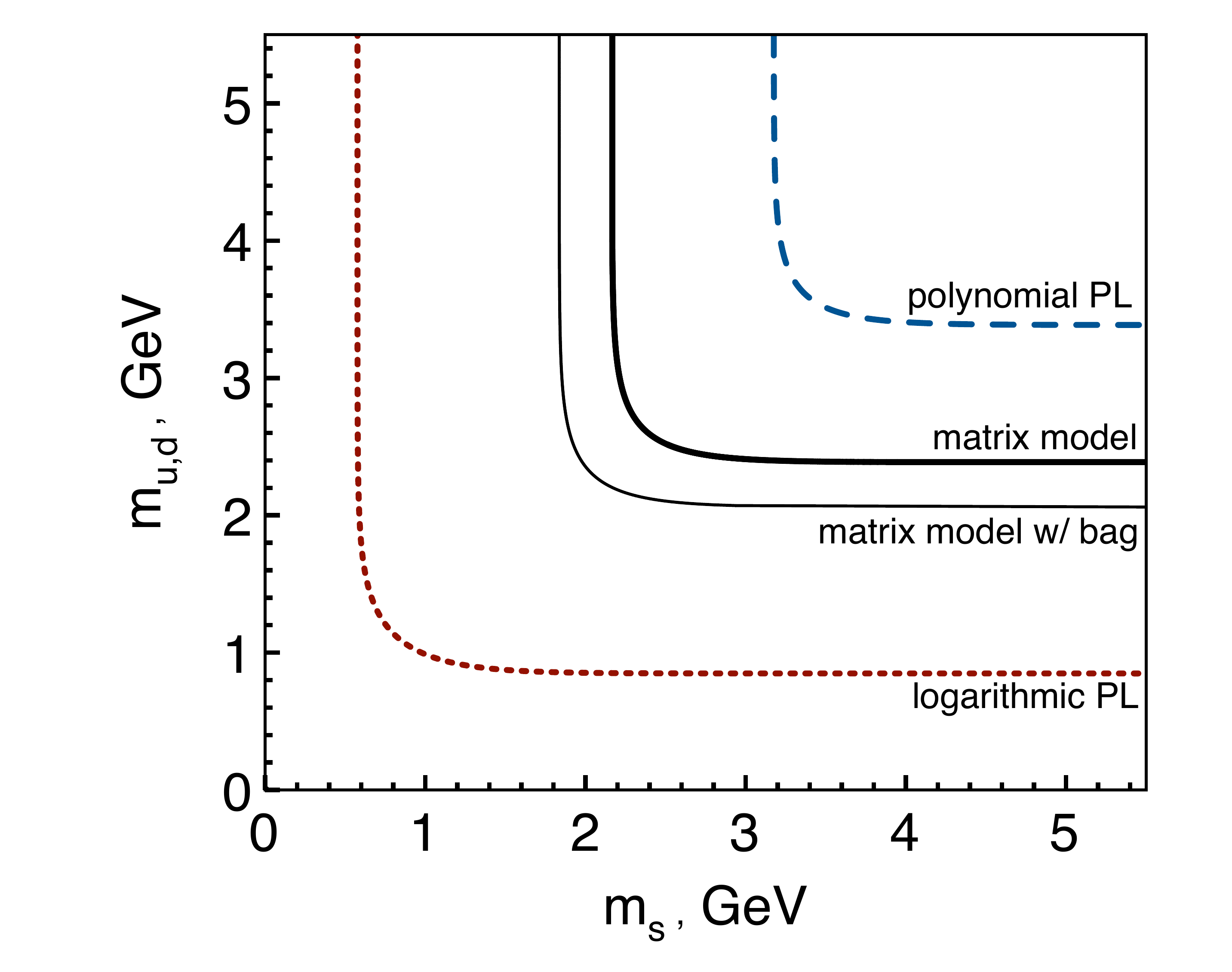}
\caption{
The phase diagram of the deconfining phase transition for the matrix model without/with bag constant (bold/thin solid line) 
and the Polyakov loop models with the polynomial (dashed line) and logarithmic (dotted line) potentials.
The lines correspond to the second order deconfining phase transition. 
}
\label{fig:TdeG}
\end{figure}

If true,
we can immediately make an interesting prediction.  When the quark
mass is heavy, the change in the critical temperature is small, and
$
\tde \approx \td
$.
For the one and two parameter matrix models
\cite{Dumitru:2010mj, Dumitru:2012},  this holds to $\sim 1\%$,
Eqs. (\ref{changetdone}) and (\ref{changetdtwo}).

Assuming that $\tde \approx \td$, 
consider a theory with $N_f$ flavors of quarks,
each with mass $m$.  The mass for the deconfining critical endpoint
is a function of $N_f$, $\mqd(N_f)$.  {\it If} the non-relativistic
approximation holds, we trivially obtain the relation
\beq
\log(N_f)
+ \frac{3}{2} \;\log\left(\frac{\mqd(N_f)}{\td} \right)
- \frac{\mqd(N_f)}{\td} \approx {\rm constant} \; .
\label{non_rel_scaling}
\eeq
Once we know $\mqd$ for one value of $N_f$, this relation
gives us its value for any $N_f$.  

In general, $\mqd$ and $\tde$ must be computed numerically.  Even
working with the exact result for the quark potential in Eq.
(\ref{quark_potential}), though, given the simplicity of the potentials
the numerical effort is minimal.  As always we assume 
that $\td = 270$~GeV.  

In Fig.~(\ref{fig:TdeM})
we present the results for the one parameter matrix model, without
a bag constant.  The critical temperature changes very little from
the pure gauge theory, 
\beq
\tde^{1 \; {\rm para.}} = .995 \, \td \; .
\label{changetdone}
\eeq
The deconfining critical endpoint occurs for a {\it very} heavy mass:
for one flavor, $\mqd  = 2167$~MeV.  The values for other 
flavors obey Eq. (\ref{non_rel_scaling}); as $N_f$ increases, so
does $\mqd$.

Going to a two parameter matrix model with a bag constant, we find that
again the shift in the critical temperature is very small,
\beq
\tde^{2 \; {\rm para.}} = .990 \, \td \; .
\label{changetdtwo}
\eeq
The quark mass $\mqd$ moves down by about $10\%$ from the
the one parameter model, with $\mqd = 1836$~MeV for one flavor.
This is the direction expected: in the one parameter 
model \cite{Dumitru:2010mj}
the latent heat is too small.  Going to a two parameter model, with a 
bag constant, increases the latent heat to agree with the 
lattice data \cite{Dumitru:2012}.  If the latent heat is larger, it
should take a larger background field, or a smaller quark mass, to wash
out the first order deconfining phase transition.  

For the zero parameter matrix model,
for one flavor we find a slightly larger shift
in the critical temperature, $\tde = 0.973 \, \td$, and a smaller
quark mass, $\mqd = 1536$~MeV.  These values
agree with those of Ref. \cite{Meisinger:2001cq}.  
This quark mass is $\sim 17\%$ lighter than for the
two parameter model, but remember that this model does not describe the
interaction measure of the pure glue theory.

We compare the results for different models in Fig.~(\ref{fig:TdeG}).
For the logarithmic
Polyakov loop model, the masses are light, $\mqd \sim 1$~GeV.  The
temperature for the deconfining critical endpoint is significantly
less than for the pure glue theory,
\beq
\tde^{{\rm Log. PLM}} = 0.90 \; \td \;. 
\eeq

A polynomial Polyakov loop model gives a very large mass,
$\mqd \sim 3.5$ GeV.  The critical temperature is very close to the pure
glue theory, $\tde \sim 0.996\; \td$.

We comment that Ref.~\cite{Dumitru:2003cf} used a polynomial Polyakov loop
model, but finds a result rather different from ours.  
Crucially, we are in accord on the value
of the background field, $h$, at which the deconfining
critical end point occurs.  We differ in how to relate this
background field to the quark mass.  
We assume that in a Polyakov loop model, that the quark 
contribution is related following 
the one loop quark determinant, Sec. (\ref{sec:quark}), which gives
$h \sim m_{qk}^{3/2} \exp(-m_{qk}/T)$
when $m_{qk} \gg T$, Eq. (\ref{quark_potential_loop}).
Ref.~\cite{Dumitru:2003cf} uses a relation motivated by light quarks,
and take $h \sim \exp( - m_{\pi})$; 
doing so then gives $m_{\pi}^{de} \sim 1.8$~GeV.

\begin{figure}
\includegraphics[scale=0.3]{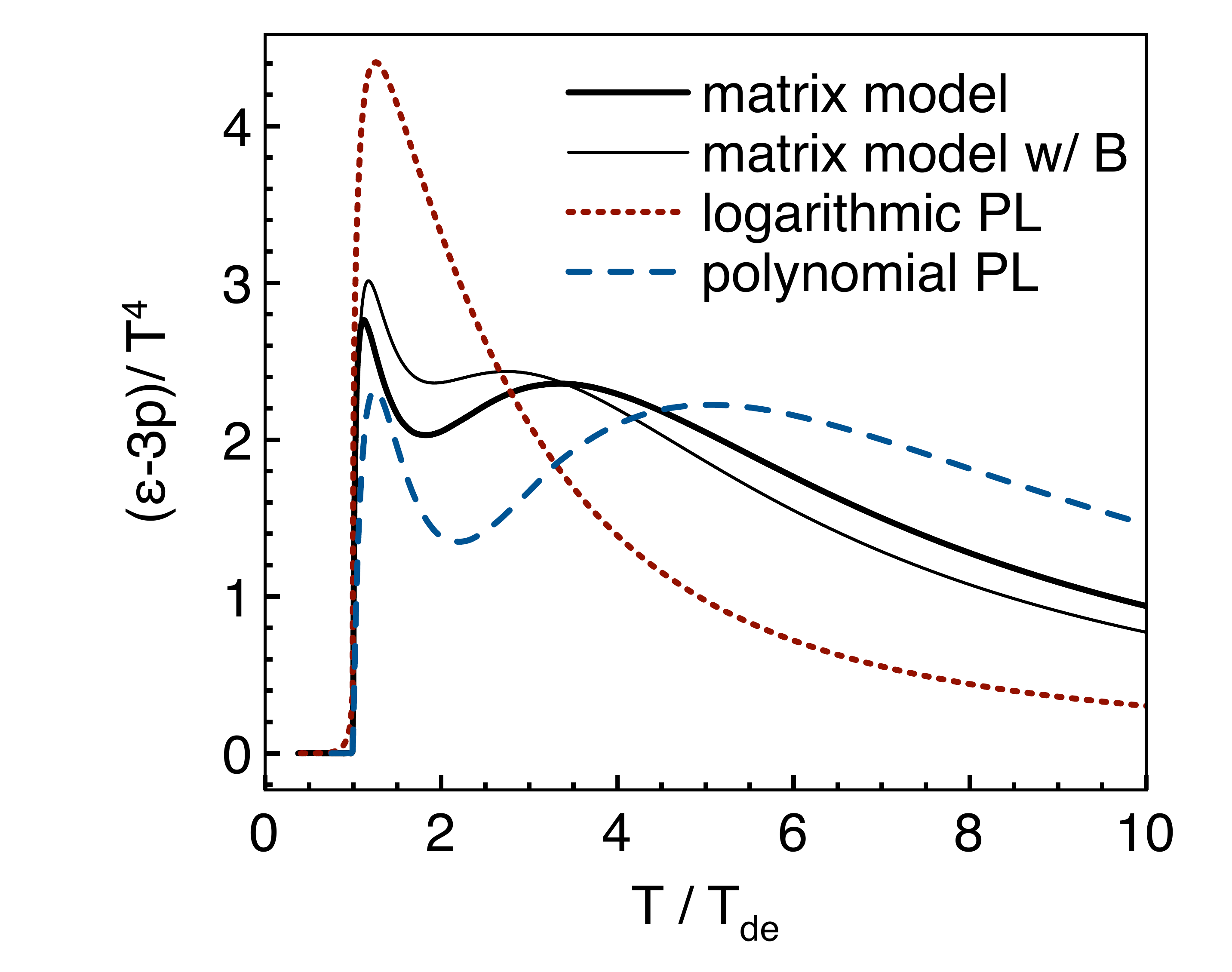}
\caption{
The conformal anomaly, $(\varepsilon-3p)/T^4$, as a function of the temperature, $T$, normalized by $T_{de}$ 
for the matrix  model without/with bag constant (bold/thin solid line) 
and the Polyakov loop models with the polynomial (dashed line) and logarithmic (dotted line) potentials.
The calculations are performed for three flavors with  $m=m^{de}$. 
}
\label{fig:CA}
\end{figure}

It is also interesting to compute the interaction measure at
the deconfining critical endpoint.  In our model, it is a sum of the
interaction measure for the glue part of the theory, plus the interaction
measure for quarks.  For a massive quark, 
it is easy to see that the interaction measure has
a peak at $\sim 0.4\;  m$.  For the logarithmic Polyakov loop potential,
because the quark mass is relatively light, $\mqd \sim 1$~GeV,
the contribution of three degenerate
flavors of quarks to the interaction measure just broadens
and enhances the peak in the pure glue theory, Fig.~(\ref{fig:CA}).  

In contrast, for either matrix model, or for a polynomial Polyakov loop
potential, the quark mass is heavier than $2$~GeV.  This means that
the interaction measure has a characteristic form: there is the usual
peak from gluons, near $\tde$, {\it plus} a second peak from quarks,
at a heavier temperature.  

This two peak structure in the interaction measure is special to three
degenerate flavors of quarks.  For one or two flavors, there is a peak
in the interaction measure from quarks, but it does not stick out over
the contribution from gluons.

\section{Conclusions}

The existence and
computation of the deconfining critical endpoint is a well known problem.
In this paper we have shown that its properties can be used to
differentiate between different effective models.

If the critical quark mass is very heavy, as in matrix and polynomial
Polyakov loop models, then effects of lattice discretization for such
heavy quarks will be severe.  It 
may be useful instead to compute the
background field at which the 
deconfining critical endpoint occurs, e.~g. by adding a term proportional to  
$h \; ( {\rm tr} \; {\bf L} + {\rm tr} \; {\bf L}^\dagger)/2$ 
to the Yang Mills action, where $h$ is the external field.
This is easily computed to one loop order
in the Boltzmann approximation.  This result receives
corrections perturbatively, proportional to $\sim g^2$, {\it etc.}.  
Nevertheless,
computing with a background field is elementary in numerical simulations
on the lattice, and should indicate if $\mqd$ is relatively light,
$\sim 1$~GeV, or heavy, $\sim 2$~GeV.

\begin{acknowledgments}
We thank A.~Bazavov, F.~Karsch, 
Y.~Maezawa,  S.~Mukherjee and, especially, 
A. Dumitru and P.~Petreczky for discussions.
The research of R.D.P. and V.S. is supported
by the U.S. Department of Energy under contract \#DE-AC02-98CH10886.
K.K. is supported by the RIKEN Special Postdoctoral Researchers Program.
\end{acknowledgments}


%merlin.mbs 2010-03-15 4.21a (PWD, AO, DPC)
\begin{thebibliography}{0}%
\makeatletter
\providecommand \@ifxundefined [1]{%
 \@ifx{#1\undefined}
}%
\providecommand \@ifnum [1]{%
 \ifnum #1\expandafter \@firstoftwo
 \else \expandafter \@secondoftwo
 \fi
}%
\providecommand \@ifx [1]{%
 \ifx #1\expandafter \@firstoftwo
 \else \expandafter \@secondoftwo
 \fi
}%
\providecommand \natexlab [1]{#1}%
\providecommand \enquote  [1]{``#1''}%
\providecommand \bibnamefont  [1]{#1}%
\providecommand \bibfnamefont [1]{#1}%
\providecommand \citenamefont [1]{#1}%
\providecommand \href@noop [0]{\@secondoftwo}%
\providecommand \href [0]{\begingroup \@sanitize@url \@href}%
\providecommand \@href[1]{\@@startlink{#1}\@@href}%
\providecommand \@@href[1]{\endgroup#1\@@endlink}%
\providecommand \@sanitize@url [0]{\catcode `\\12\catcode `\$12\catcode
  `\&12\catcode `\#12\catcode `\^12\catcode `\_12\catcode `\%12\relax}%
\providecommand \@@startlink[1]{}%
\providecommand \@@endlink[0]{}%
\providecommand \url  [0]{\begingroup\@sanitize@url \@url }%
\providecommand \@url [1]{\endgroup\@href {#1}{\urlprefix }}%
\providecommand \urlprefix  [0]{URL }%
\providecommand \Eprint [0]{\href }%
\@ifxundefined \urlstyle {%
  \providecommand \doi  [0]{\begingroup \@sanitize@url \@doi}%
  \providecommand \@doi [1]{\endgroup \@@startlink {\doibase
  #1}doi:\discretionary {}{}{}#1\@@endlink }%
}{%
  \providecommand \doi  [0]{doi:\discretionary{}{}{}\begingroup
  \urlstyle{rm}\Url }%
}%
\providecommand \doibase [0]{http://dx.doi.org/}%
\providecommand \Doi [0]{\begingroup \@sanitize@url \@Doi }%
\providecommand \@Doi  [1]{\endgroup\@@startlink{\doibase#1}\@@Doi}%
\providecommand \@@Doi [1]{#1\@@endlink}%
\providecommand \selectlanguage [0]{\@gobble}%
\providecommand \bibinfo  [0]{\@secondoftwo}%
\providecommand \bibfield  [0]{\@secondoftwo}%
\providecommand \translation [1]{[#1]}%
\providecommand \BibitemOpen [0]{}%
\providecommand \bibitemStop [0]{}%
\providecommand \bibitemNoStop [0]{.\EOS\space}%
\providecommand \EOS [0]{\spacefactor3000\relax}%
\providecommand \BibitemShut  [1]{\csname bibitem#1\endcsname}%
%</preamble>
\end{thebibliography}%


%merlin.mbs 2010-03-15 4.21a (PWD, AO, DPC)
\begin{thebibliography}{24}%
\makeatletter
\providecommand \@ifxundefined [1]{%
 \@ifx{#1\undefined}
}%
\providecommand \@ifnum [1]{%
 \ifnum #1\expandafter \@firstoftwo
 \else \expandafter \@secondoftwo
 \fi
}%
\providecommand \@ifx [1]{%
 \ifx #1\expandafter \@firstoftwo
 \else \expandafter \@secondoftwo
 \fi
}%
\providecommand \natexlab [1]{#1}%
\providecommand \enquote  [1]{``#1''}%
\providecommand \bibnamefont  [1]{#1}%
\providecommand \bibfnamefont [1]{#1}%
\providecommand \citenamefont [1]{#1}%
\providecommand \href@noop [0]{\@secondoftwo}%
\providecommand \href [0]{\begingroup \@sanitize@url \@href}%
\providecommand \@href[1]{\@@startlink{#1}\@@href}%
\providecommand \@@href[1]{\endgroup#1\@@endlink}%
\providecommand \@sanitize@url [0]{\catcode `\\12\catcode `\$12\catcode
  `\&12\catcode `\#12\catcode `\^12\catcode `\_12\catcode `\%12\relax}%
\providecommand \@@startlink[1]{}%
\providecommand \@@endlink[0]{}%
\providecommand \url  [0]{\begingroup\@sanitize@url \@url }%
\providecommand \@url [1]{\endgroup\@href {#1}{\urlprefix }}%
\providecommand \urlprefix  [0]{URL }%
\providecommand \Eprint [0]{\href }%
\@ifxundefined \urlstyle {%
  \providecommand \doi  [0]{\begingroup \@sanitize@url \@doi}%
  \providecommand \@doi [1]{\endgroup \@@startlink {\doibase
  #1}doi:\discretionary {}{}{}#1\@@endlink }%
}{%
  \providecommand \doi  [0]{doi:\discretionary{}{}{}\begingroup
  \urlstyle{rm}\Url }%
}%
\providecommand \doibase [0]{http://dx.doi.org/}%
\providecommand \Doi [0]{\begingroup \@sanitize@url \@Doi }%
\providecommand \@Doi  [1]{\endgroup\@@startlink{\doibase#1}\@@Doi}%
\providecommand \@@Doi [1]{#1\@@endlink}%
\providecommand \selectlanguage [0]{\@gobble}%
\providecommand \bibinfo  [0]{\@secondoftwo}%
\providecommand \bibfield  [0]{\@secondoftwo}%
\providecommand \translation [1]{[#1]}%
\providecommand \BibitemOpen [0]{}%
\providecommand \bibitemStop [0]{}%
\providecommand \bibitemNoStop [0]{.\EOS\space}%
\providecommand \EOS [0]{\spacefactor3000\relax}%
\providecommand \BibitemShut  [1]{\csname bibitem#1\endcsname}%
%</preamble>
\bibitem [{\citenamefont {Svetitsky}\ and\ \citenamefont
  {Yaffe}(1982)}]{Svetitsky:1982gs}%
  \BibitemOpen
  \bibfield  {author} {\bibinfo {author} {\bibfnamefont {B.}~\bibnamefont
  {Svetitsky}}\ and\ \bibinfo {author} {\bibfnamefont {L.~G.}\ \bibnamefont
  {Yaffe}},\ }\Doi {10.1016/0550-3213(82)90172-9} {\bibfield  {journal}
  {\bibinfo  {journal} {Nucl.Phys.},\ }\textbf {\bibinfo {volume} {B210}},\
  \bibinfo {pages} {423} (\bibinfo {year} {1982})}\BibitemShut {NoStop}%
\bibitem [{\citenamefont {Meisinger}\ \emph {et~al.}(2002)\citenamefont
  {Meisinger}, \citenamefont {Miller},\ and\ \citenamefont
  {Ogilvie}}]{Meisinger:2001cq}%
  \BibitemOpen
  \bibfield  {author} {\bibinfo {author} {\bibfnamefont {P.~N.}\ \bibnamefont
  {Meisinger}}, \bibinfo {author} {\bibfnamefont {T.~R.}\ \bibnamefont
  {Miller}}, \ and\ \bibinfo {author} {\bibfnamefont {M.~C.}\ \bibnamefont
  {Ogilvie}},\ }\Doi {10.1103/PhysRevD.65.034009} {\bibfield  {journal}
  {\bibinfo  {journal} {Phys.Rev.},\ }\textbf {\bibinfo {volume} {D65}},\
  \bibinfo {pages} {034009} (\bibinfo {year} {2002})},\ \Eprint
  {http://arxiv.org/abs/hep-ph/0108009} {arXiv:hep-ph/0108009 [hep-ph]}
  \BibitemShut {NoStop}%
\bibitem [{\citenamefont {Meisinger}\ and\ \citenamefont
  {Ogilvie}(2002)}]{Meisinger:2001fi}%
  \BibitemOpen
  \bibfield  {author} {\bibinfo {author} {\bibfnamefont {P.~N.}\ \bibnamefont
  {Meisinger}}\ and\ \bibinfo {author} {\bibfnamefont {M.~C.}\ \bibnamefont
  {Ogilvie}},\ }\Doi {10.1103/PhysRevD.65.056013} {\bibfield  {journal}
  {\bibinfo  {journal} {Phys.Rev.},\ }\textbf {\bibinfo {volume} {D65}},\
  \bibinfo {pages} {056013} (\bibinfo {year} {2002})},\ \Eprint
  {http://arxiv.org/abs/hep-ph/0108026} {arXiv:hep-ph/0108026 [hep-ph]}
  \BibitemShut {NoStop}%
%%CITATION = HEP-PH/0108026;%%
\bibitem [{\citenamefont {Dumitru}\ \emph {et~al.}(2011)\citenamefont
  {Dumitru}, \citenamefont {Guo}, \citenamefont {Hidaka}, \citenamefont
  {Korthals~Altes},\ and\ \citenamefont {Pisarski}}]{Dumitru:2010mj}%
  \BibitemOpen
  \bibfield  {author} {\bibinfo {author} {\bibfnamefont {A.}~\bibnamefont
  {Dumitru}}, \bibinfo {author} {\bibfnamefont {Y.}~\bibnamefont {Guo}},
  \bibinfo {author} {\bibfnamefont {Y.}~\bibnamefont {Hidaka}}, \bibinfo
  {author} {\bibfnamefont {C.~P.}\ \bibnamefont {Korthals~Altes}}, \ and\
  \bibinfo {author} {\bibfnamefont {R.~D.}\ \bibnamefont {Pisarski}},\ }\Doi
  {10.1103/PhysRevD.83.034022} {\bibfield  {journal} {\bibinfo  {journal}
  {Phys.Rev.},\ }\textbf {\bibinfo {volume} {D83}},\ \bibinfo {pages} {034022}
  (\bibinfo {year} {2011})},\ \Eprint {http://arxiv.org/abs/1011.3820}
  {arXiv:1011.3820 [hep-ph]} \BibitemShut {NoStop}%
\bibitem [{\citenamefont {Dumitru}\ \emph {et~al.}(2012)\citenamefont
  {Dumitru}, \citenamefont {Guo}, \citenamefont {Hidaka}, \citenamefont
  {Korthals~Altes},\ and\ \citenamefont {Pisarski}}]{Dumitru:2012}%
  \BibitemOpen
  \bibfield  {author} {\bibinfo {author} {\bibfnamefont {A.}~\bibnamefont
  {Dumitru}}, \bibinfo {author} {\bibfnamefont {Y.}~\bibnamefont {Guo}},
  \bibinfo {author} {\bibfnamefont {Y.}~\bibnamefont {Hidaka}}, \bibinfo
  {author} {\bibfnamefont {C.~P.}\ \bibnamefont {Korthals~Altes}}, \ and\
  \bibinfo {author} {\bibfnamefont {R.~D.}\ \bibnamefont {Pisarski}},\
  }\href@noop {} {\enquote {\bibinfo {title} {{Effective matrix models for
  deconfinement in pure glue theories}},}\ } (\bibinfo {year}
  {2012})\BibitemShut {NoStop}%
\bibitem [{\citenamefont {Pisarski}(2000)}]{Pisarski:2000eq}%
  \BibitemOpen
  \bibfield  {author} {\bibinfo {author} {\bibfnamefont {R.~D.}\ \bibnamefont
  {Pisarski}},\ }\Doi {10.1103/PhysRevD.62.111501} {\bibfield  {journal}
  {\bibinfo  {journal} {Phys.Rev.},\ }\textbf {\bibinfo {volume} {D62}},\
  \bibinfo {pages} {111501} (\bibinfo {year} {2000})},\ \Eprint
  {http://arxiv.org/abs/hep-ph/0006205} {arXiv:hep-ph/0006205 [hep-ph]}
  \BibitemShut {NoStop}%
%%CITATION = HEP-PH/0006205;%%
\bibitem [{\citenamefont {Scavenius}\ \emph {et~al.}(2002)\citenamefont
  {Scavenius}, \citenamefont {Dumitru},\ and\ \citenamefont
  {Lenaghan}}]{Scavenius:2002ru}%
  \BibitemOpen
  \bibfield  {author} {\bibinfo {author} {\bibfnamefont {O.}~\bibnamefont
  {Scavenius}}, \bibinfo {author} {\bibfnamefont {A.}~\bibnamefont {Dumitru}},
  \ and\ \bibinfo {author} {\bibfnamefont {J.}~\bibnamefont {Lenaghan}},\ }\Doi
  {10.1103/PhysRevC.66.034903} {\bibfield  {journal} {\bibinfo  {journal}
  {Phys.Rev.},\ }\textbf {\bibinfo {volume} {C66}},\ \bibinfo {pages} {034903}
  (\bibinfo {year} {2002})},\ \Eprint {http://arxiv.org/abs/hep-ph/0201079}
  {arXiv:hep-ph/0201079 [hep-ph]} \BibitemShut {NoStop}%
%%CITATION = HEP-PH/0201079;%%
\bibitem [{\citenamefont {Dumitru}\ \emph {et~al.}(2004)\citenamefont
  {Dumitru}, \citenamefont {Roder},\ and\ \citenamefont
  {Ruppert}}]{Dumitru:2003cf}%
  \BibitemOpen
  \bibfield  {author} {\bibinfo {author} {\bibfnamefont {A.}~\bibnamefont
  {Dumitru}}, \bibinfo {author} {\bibfnamefont {D.}~\bibnamefont {Roder}}, \
  and\ \bibinfo {author} {\bibfnamefont {J.}~\bibnamefont {Ruppert}},\ }\Doi
  {10.1103/PhysRevD.70.074001} {\bibfield  {journal} {\bibinfo  {journal}
  {Phys.Rev.},\ }\textbf {\bibinfo {volume} {D70}},\ \bibinfo {pages} {074001}
  (\bibinfo {year} {2004})},\ \Eprint {http://arxiv.org/abs/hep-ph/0311119}
  {arXiv:hep-ph/0311119 [hep-ph]} \BibitemShut {NoStop}%
%%CITATION = HEP-PH/0311119;%%
\bibitem [{\citenamefont {Ratti}\ \emph {et~al.}(2006)\citenamefont {Ratti},
  \citenamefont {Thaler},\ and\ \citenamefont {Weise}}]{Ratti:2005jh}%
  \BibitemOpen
  \bibfield  {author} {\bibinfo {author} {\bibfnamefont {C.}~\bibnamefont
  {Ratti}}, \bibinfo {author} {\bibfnamefont {M.~A.}\ \bibnamefont {Thaler}}, \
  and\ \bibinfo {author} {\bibfnamefont {W.}~\bibnamefont {Weise}},\ }\Doi
  {10.1103/PhysRevD.73.014019} {\bibfield  {journal} {\bibinfo  {journal}
  {Phys.Rev.},\ }\textbf {\bibinfo {volume} {D73}},\ \bibinfo {pages} {014019}
  (\bibinfo {year} {2006})},\ \Eprint {http://arxiv.org/abs/hep-ph/0506234}
  {arXiv:hep-ph/0506234 [hep-ph]} \BibitemShut {NoStop}%
%%CITATION = HEP-PH/0506234;%%
\bibitem [{\citenamefont {Fukushima}(2004)}]{Fukushima:2003fw}%
  \BibitemOpen
  \bibfield  {author} {\bibinfo {author} {\bibfnamefont {K.}~\bibnamefont
  {Fukushima}},\ }\Doi {10.1016/j.physletb.2004.04.027} {\bibfield  {journal}
  {\bibinfo  {journal} {Phys.Lett.},\ }\textbf {\bibinfo {volume} {B591}},\
  \bibinfo {pages} {277} (\bibinfo {year} {2004})},\ \Eprint
  {http://arxiv.org/abs/hep-ph/0310121} {arXiv:hep-ph/0310121 [hep-ph]}
  \BibitemShut {NoStop}%
%%CITATION = HEP-PH/0310121;%%
\bibitem [{\citenamefont {Roessner}\ \emph {et~al.}(2007)\citenamefont
  {Roessner}, \citenamefont {Ratti},\ and\ \citenamefont
  {Weise}}]{Roessner:2006xn}%
  \BibitemOpen
  \bibfield  {author} {\bibinfo {author} {\bibfnamefont {S.}~\bibnamefont
  {Roessner}}, \bibinfo {author} {\bibfnamefont {C.}~\bibnamefont {Ratti}}, \
  and\ \bibinfo {author} {\bibfnamefont {W.}~\bibnamefont {Weise}},\ }\Doi
  {10.1103/PhysRevD.75.034007} {\bibfield  {journal} {\bibinfo  {journal}
  {Phys.Rev.},\ }\textbf {\bibinfo {volume} {D75}},\ \bibinfo {pages} {034007}
  (\bibinfo {year} {2007})},\ \Eprint {http://arxiv.org/abs/hep-ph/0609281}
  {arXiv:hep-ph/0609281 [hep-ph]} \BibitemShut {NoStop}%
%%CITATION = HEP-PH/0609281;%%
\bibitem [{\citenamefont {Hell}\ \emph {et~al.}(2010)\citenamefont {Hell},
  \citenamefont {Rossner}, \citenamefont {Cristoforetti},\ and\ \citenamefont
  {Weise}}]{Hell:2009by}%
  \BibitemOpen
  \bibfield  {author} {\bibinfo {author} {\bibfnamefont {T.}~\bibnamefont
  {Hell}}, \bibinfo {author} {\bibfnamefont {S.}~\bibnamefont {Rossner}},
  \bibinfo {author} {\bibfnamefont {M.}~\bibnamefont {Cristoforetti}}, \ and\
  \bibinfo {author} {\bibfnamefont {W.}~\bibnamefont {Weise}},\ }\Doi
  {10.1103/PhysRevD.81.074034} {\bibfield  {journal} {\bibinfo  {journal}
  {Phys.Rev.},\ }\textbf {\bibinfo {volume} {D81}},\ \bibinfo {pages} {074034}
  (\bibinfo {year} {2010})},\ \Eprint {http://arxiv.org/abs/0911.3510}
  {arXiv:0911.3510 [hep-ph]} \BibitemShut {NoStop}%
%%CITATION = ARXIV:0911.3510;%%
\bibitem [{\citenamefont {Horvatic}\ \emph {et~al.}(2011)\citenamefont
  {Horvatic}, \citenamefont {Blaschke}, \citenamefont {Klabucar},\ and\
  \citenamefont {Kaczmarek}}]{Horvatic:2010md}%
  \BibitemOpen
  \bibfield  {author} {\bibinfo {author} {\bibfnamefont {D.}~\bibnamefont
  {Horvatic}}, \bibinfo {author} {\bibfnamefont {D.}~\bibnamefont {Blaschke}},
  \bibinfo {author} {\bibfnamefont {D.}~\bibnamefont {Klabucar}}, \ and\
  \bibinfo {author} {\bibfnamefont {O.}~\bibnamefont {Kaczmarek}},\ }\Doi
  {10.1103/PhysRevD.84.016005} {\bibfield  {journal} {\bibinfo  {journal}
  {Phys.Rev.},\ }\textbf {\bibinfo {volume} {D84}},\ \bibinfo {pages} {016005}
  (\bibinfo {year} {2011})},\ \Eprint {http://arxiv.org/abs/1012.2113}
  {arXiv:1012.2113 [hep-ph]} \BibitemShut {NoStop}%
%%CITATION = ARXIV:1012.2113;%%
\bibitem [{\citenamefont {Hell}\ \emph {et~al.}(2011)\citenamefont {Hell},
  \citenamefont {Kashiwa},\ and\ \citenamefont {Weise}}]{Hell:2011ic}%
  \BibitemOpen
  \bibfield  {author} {\bibinfo {author} {\bibfnamefont {T.}~\bibnamefont
  {Hell}}, \bibinfo {author} {\bibfnamefont {K.}~\bibnamefont {Kashiwa}}, \
  and\ \bibinfo {author} {\bibfnamefont {W.}~\bibnamefont {Weise}},\ }\Doi
  {10.1103/PhysRevD.83.114008} {\bibfield  {journal} {\bibinfo  {journal}
  {Phys.Rev.},\ }\textbf {\bibinfo {volume} {D83}},\ \bibinfo {pages} {114008}
  (\bibinfo {year} {2011})},\ \Eprint {http://arxiv.org/abs/1104.0572}
  {arXiv:1104.0572 [hep-ph]} \BibitemShut {NoStop}%
%%CITATION = ARXIV:1104.0572;%%
\bibitem [{\citenamefont {DeTar}\ and\ \citenamefont
  {Heller}(2009)}]{DeTar:2009ef}%
  \BibitemOpen
  \bibfield  {author} {\bibinfo {author} {\bibfnamefont {C.}~\bibnamefont
  {DeTar}}\ and\ \bibinfo {author} {\bibfnamefont {U.}~\bibnamefont {Heller}},\
  }\Doi {10.1140/epja/i2009-10825-3} {\bibfield  {journal} {\bibinfo  {journal}
  {Eur.Phys.J.},\ }\textbf {\bibinfo {volume} {A41}},\ \bibinfo {pages} {405}
  (\bibinfo {year} {2009})},\ \Eprint {http://arxiv.org/abs/0905.2949}
  {arXiv:0905.2949 [hep-lat]} \BibitemShut {NoStop}%
%%CITATION = ARXIV:0905.2949;%%
\bibitem [{\citenamefont {Petreczky}(2012)}]{Petreczky:2012rq}%
  \BibitemOpen
  \bibfield  {author} {\bibinfo {author} {\bibfnamefont {P.}~\bibnamefont
  {Petreczky}},\ }\href@noop {} { (\bibinfo {year} {2012})},\ \Eprint
  {http://arxiv.org/abs/1203.5320} {arXiv:1203.5320 [hep-lat]} \BibitemShut
  {NoStop}%
%%CITATION = ARXIV:1203.5320;%%
\bibitem [{\citenamefont {Meisinger}\ and\ \citenamefont
  {Ogilvie}(1995)}]{Meisinger:1995qr}%
  \BibitemOpen
  \bibfield  {author} {\bibinfo {author} {\bibfnamefont {P.~N.}\ \bibnamefont
  {Meisinger}}\ and\ \bibinfo {author} {\bibfnamefont {M.~C.}\ \bibnamefont
  {Ogilvie}},\ }\Doi {10.1103/PhysRevD.52.3024} {\bibfield  {journal} {\bibinfo
   {journal} {Phys.Rev.},\ }\textbf {\bibinfo {volume} {D52}},\ \bibinfo
  {pages} {3024} (\bibinfo {year} {1995})},\ \Eprint
  {http://arxiv.org/abs/hep-lat/9502003} {arXiv:hep-lat/9502003 [hep-lat]}
  \BibitemShut {NoStop}%
%%CITATION = HEP-LAT/9502003;%%
\bibitem [{\citenamefont {Alexandrou}\ \emph {et~al.}(1999)\citenamefont
  {Alexandrou}, \citenamefont {Borici}, \citenamefont {Feo}, \citenamefont
  {de~Forcrand}, \citenamefont {Galli} \emph {et~al.}}]{Alexandrou:1998wv}%
  \BibitemOpen
  \bibfield  {author} {\bibinfo {author} {\bibfnamefont {C.}~\bibnamefont
  {Alexandrou}}, \bibinfo {author} {\bibfnamefont {A.}~\bibnamefont {Borici}},
  \bibinfo {author} {\bibfnamefont {A.}~\bibnamefont {Feo}}, \bibinfo {author}
  {\bibfnamefont {P.}~\bibnamefont {de~Forcrand}}, \bibinfo {author}
  {\bibfnamefont {A.}~\bibnamefont {Galli}},  \emph {et~al.},\ }\Doi
  {10.1103/PhysRevD.60.034504} {\bibfield  {journal} {\bibinfo  {journal}
  {Phys.Rev.},\ }\textbf {\bibinfo {volume} {D60}},\ \bibinfo {pages} {034504}
  (\bibinfo {year} {1999})},\ \Eprint {http://arxiv.org/abs/hep-lat/9811028}
  {arXiv:hep-lat/9811028 [hep-lat]} \BibitemShut {NoStop}%
%%CITATION = HEP-LAT/9811028;%%
\bibitem [{\citenamefont {Karsch}\ \emph {et~al.}(2002)\citenamefont {Karsch},
  \citenamefont {Schmidt},\ and\ \citenamefont {Stickan}}]{Karsch:2001ya}%
  \BibitemOpen
  \bibfield  {author} {\bibinfo {author} {\bibfnamefont {F.}~\bibnamefont
  {Karsch}}, \bibinfo {author} {\bibfnamefont {C.}~\bibnamefont {Schmidt}}, \
  and\ \bibinfo {author} {\bibfnamefont {S.}~\bibnamefont {Stickan}},\ }\Doi
  {10.1016/S0010-4655(02)00327-2} {\bibfield  {journal} {\bibinfo  {journal}
  {Comput.Phys.Commun.},\ }\textbf {\bibinfo {volume} {147}},\ \bibinfo {pages}
  {451} (\bibinfo {year} {2002})},\ \Eprint
  {http://arxiv.org/abs/hep-lat/0111059} {arXiv:hep-lat/0111059 [hep-lat]}
  \BibitemShut {NoStop}%
%%CITATION = HEP-LAT/0111059;%%
\bibitem [{\citenamefont {Saito}\ \emph {et~al.}(2011)\citenamefont {Saito}
  \emph {et~al.}}]{Saito:2011fs}%
  \BibitemOpen
  \bibfield  {author} {\bibinfo {author} {\bibfnamefont {H.}~\bibnamefont
  {Saito}} \emph {et~al.} (\bibinfo {collaboration} {WHOT-QCD Collaboration}),\
  }\Doi {10.1103/PhysRevD.84.054502} {\bibfield  {journal} {\bibinfo  {journal}
  {Phys.Rev.},\ }\textbf {\bibinfo {volume} {D84}},\ \bibinfo {pages} {054502}
  (\bibinfo {year} {2011})},\ \Eprint {http://arxiv.org/abs/1106.0974}
  {arXiv:1106.0974 [hep-lat]} \BibitemShut {NoStop}%
%%CITATION = ARXIV:1106.0974;%%
\bibitem [{\citenamefont {Borsanyi}\ \emph {et~al.}(2012)\citenamefont
  {Borsanyi}, \citenamefont {Endrodi}, \citenamefont {Fodor}, \citenamefont
  {Katz}, \citenamefont {Krieg} \emph {et~al.}}]{Borsanyi:2012vn}%
  \BibitemOpen
  \bibfield  {author} {\bibinfo {author} {\bibfnamefont {S.}~\bibnamefont
  {Borsanyi}}, \bibinfo {author} {\bibfnamefont {G.}~\bibnamefont {Endrodi}},
  \bibinfo {author} {\bibfnamefont {Z.}~\bibnamefont {Fodor}}, \bibinfo
  {author} {\bibfnamefont {S.~D.}\ \bibnamefont {Katz}}, \bibinfo {author}
  {\bibfnamefont {S.}~\bibnamefont {Krieg}},  \emph {et~al.},\ }\href@noop {} {
  (\bibinfo {year} {2012})},\ \Eprint {http://arxiv.org/abs/1204.0995}
  {arXiv:1204.0995 [hep-lat]} \BibitemShut {NoStop}%
%%CITATION = ARXIV:1204.0995;%%
\bibitem [{\citenamefont {Pisarski}\ and\ \citenamefont
  {Skokov}()}]{Pisarski:2012}%
  \BibitemOpen
  \bibfield  {author} {\bibinfo {author} {\bibfnamefont {R.~D.}\ \bibnamefont
  {Pisarski}}\ and\ \bibinfo {author} {\bibfnamefont {V.~V.}\ \bibnamefont
  {Skokov}},\ }\href@noop {} {\enquote {\bibinfo {title} {{unpublished}},}\
  }\BibitemShut {NoStop}%
\bibitem [{\citenamefont {Boyd}\ \emph {et~al.}(1996)\citenamefont {Boyd},
  \citenamefont {Engels}, \citenamefont {Karsch}, \citenamefont {Laermann},
  \citenamefont {Legeland} \emph {et~al.}}]{Boyd:1996bx}%
  \BibitemOpen
  \bibfield  {author} {\bibinfo {author} {\bibfnamefont {G.}~\bibnamefont
  {Boyd}}, \bibinfo {author} {\bibfnamefont {J.}~\bibnamefont {Engels}},
  \bibinfo {author} {\bibfnamefont {F.}~\bibnamefont {Karsch}}, \bibinfo
  {author} {\bibfnamefont {E.}~\bibnamefont {Laermann}}, \bibinfo {author}
  {\bibfnamefont {C.}~\bibnamefont {Legeland}},  \emph {et~al.},\ }\Doi
  {10.1016/0550-3213(96)00170-8} {\bibfield  {journal} {\bibinfo  {journal}
  {Nucl.Phys.},\ }\textbf {\bibinfo {volume} {B469}},\ \bibinfo {pages} {419}
  (\bibinfo {year} {1996})},\ \Eprint {http://arxiv.org/abs/hep-lat/9602007}
  {arXiv:hep-lat/9602007 [hep-lat]} \BibitemShut {NoStop}%
%%CITATION = HEP-LAT/9602007;%%
\bibitem [{\citenamefont {Kaczmarek}\ \emph {et~al.}(2002)\citenamefont
  {Kaczmarek}, \citenamefont {Karsch}, \citenamefont {Petreczky},\ and\
  \citenamefont {Zantow}}]{Kaczmarek:2002mc}%
  \BibitemOpen
  \bibfield  {author} {\bibinfo {author} {\bibfnamefont {O.}~\bibnamefont
  {Kaczmarek}}, \bibinfo {author} {\bibfnamefont {F.}~\bibnamefont {Karsch}},
  \bibinfo {author} {\bibfnamefont {P.}~\bibnamefont {Petreczky}}, \ and\
  \bibinfo {author} {\bibfnamefont {F.}~\bibnamefont {Zantow}},\ }\Doi
  {10.1016/S0370-2693(02)02415-2} {\bibfield  {journal} {\bibinfo  {journal}
  {Phys.Lett.},\ }\textbf {\bibinfo {volume} {B543}},\ \bibinfo {pages} {41}
  (\bibinfo {year} {2002})},\ \Eprint {http://arxiv.org/abs/hep-lat/0207002}
  {arXiv:hep-lat/0207002 [hep-lat]} \BibitemShut {NoStop}%
%%CITATION = HEP-LAT/0207002;%%
\end{thebibliography}%
\end{document}